\documentclass[11pt,a4paper]{article} 
\usepackage[utf8]{inputenc} 
\usepackage{graphicx} 
\usepackage{color} 
\usepackage{fullpage}

\usepackage{comment}

\usepackage{multirow}

\usepackage{amsmath}
\usepackage{amssymb}
\usepackage{amsfonts}

\usepackage{longtable}

\usepackage{booktabs} 
\usepackage{multicol}
\usepackage{tabularx}
\usepackage{slashbox}
\usepackage{threeparttable}
\usepackage{colortbl}

\usepackage{algorithm}
\usepackage{algpseudocode}

\usepackage{optidef}
\usepackage{tikz}
\usepackage{authblk}

\usepackage{soul}

\providecommand{\keywords}[1]
{
  \small	
  \textbf{\textit{Keywords---}} #1
}

\newcommand{\gt}{GT }
\newcommand{\gtdp}{GTDP }

\newcommand{\req}[1]{(\ref{#1})}

\title{Group design in group testing for COVID-19 :\\
A French case-study.}
\author[1]{Tifaout Almeftah}
\author[1]{Luce Brotcorne}
\author[2]{Diego Cattaruzza}
\author[3,1]{Bernard Fortz}
\author[1]{Kaba Keita}
\author[3,1]{Martine Labb\'e}
\author[2]{Maxime Ogier}
\author[2]{Fr\'ed\'eric Semet}

\affil[1]{Inria Centre de recherche Lille Nord Europe, INOCS,  F-59000 Lille, France}
\affil[2]{Universit\'e de Lille, CNRS, Centrale Lille, Inria, UMR 9189 - CRIStAL, F-59000 Lille, France}
\affil[3]{Department of Computer Science, Universit\'e libre de Bruxelles, 1050 Brussels, Belgium}

\begin{document}

\maketitle

\begin{abstract}
 Group testing is a screening strategy that involves dividing a population into several disjointed groups of subjects. In its simplest implementation, each group is tested with a single test in the first phase, while in the second phase only subjects in positive groups, if any, need to be tested again individually. In this paper, we address the problem of group testing design, which aims to determine a partition into groups of a finite population in such a way that cardinality constraints on the size of each group and a constraint on the expected total number of tests are satisfied while minimizing a linear combination of the expected number of false negative and false positive classifications. First, we show that the properties and model introduced by Aprahmian et al. can be extended to the group test design problem, which is then modeled as a constrained shortest path problem on a specific graph. We design and implement an ad hoc algorithm to solve this problem. On instances based on Sant\'e Publique France data on Covid-19 screening tests, the results of the computational experiments are very promising.
\end{abstract}

\section{Introduction}

As COVID-19 has considerably impacted public health, some countries have privileged herd (collective) immunity, whereas many countries have undergone national lockdowns, which has significantly threatened the world economically and socially. By the end of spring 2020, when the spread and mortality rate of Covid-19 appeared to be under control, several countries have lifted lockdowns while respecting various measures (e.g., social distancing, wearing masks) to revive the economy of the nation and the education of youth.

The fact that the after lockdown has known an exploding number of infected cases and disease clusters is proving to be very worrying. Indeed, a second epidemic peak may force countries to decide on a local or general containment and, consequently, have an impact on all economic and social sectors trying to recover.

It is clear that no one can be spared from the virus. A significant number of patients can carry the virus without realizing it. They may be asymptomatic or pre-symptomatic, so they naturally continue to live their normal lives and are not isolated in the meantime. An increase in the number of undetected asymptomatic persons leads to an exponential increase in coronavirus spread, making the threat more serious. Therefore, detecting massively daily Covid-19 carriers in the population is an urgent necessity to slow the uncontrollable spread of the virus.

The operation of screening a population of individuals to detect the infected subjects and the non-infected ones can rapidly consume huge resources (time, effort, money). The need for rapid test results is very important to reduce laboratory congestion, identify positive individuals, isolate them, and provide appropriate treatment when necessary. Moreover, it allows to better distinguish true negatives (who may feel some symptoms from other viral infections similar to those of Covid-19) and allow them to continue their daily social and economic activities. Therefore, massive tests are very important to screen the whole population. However, we cannot lower the classification quality, as it would let false negatives in the population and favor the spread of the virus. 

From an economic point of view, a single Real-Time Polymerase Chain Reactor (RT-PCR) test, including the two steps of collecting individuals swab specimens and using pricey sophisticated equipment, can rapidly reach an approximate cost of 74 euros for the french Social Security, which leads to 300 millions euros per month. 

Last, massive testing means more congestion in laboratories. There is a crucial need to reduce as much possible the stress felt by medical crews in laboratories and the costs of testing protocols.

Thus, reducing the cost and the social impact of classifying the population into two sets appears important. One strategy is to test samples of the population instead of testing each person individually, consequently saving scarce resources while maintaining the classification accuracy.

Dorfman initially introduced this method in 1943 \cite{dorfman1943} as Group Testing (\gt). The principle of \gt in its basic implementation is to divide the population into several disjoint groups of subjects, and, in the first stage, test each group with a single test. In the second stage, only subjects in positive groups, if any, need to be tested again individually. When the risk of being infected is low, especially with large scale populations, \gt can be a highly beneficial strategy to screen all the subjects with minimal costs. In the case of a pandemic as Covid-19, the subjects to test can be pooled in groups where their samples are mixed and then tested with only one test. If viral RNA is not detected in the mixed samples, no one in the group is infected. Otherwise, at least one subject of the group is positive. In this case, a second phase of testing is fundamental to test every subject's sample individually to screen the infected patients.

Dorfman's scheme considers several theoretical assumptions as perfect tests, an infinite population, and the homogeneity of risk in the population that may not appear in real life. Indeed, the sensibility and the sensitivity of the tests play a major role in the possibility of having subjects classified as false-negative and false-positive. In reality, as the probability of being infected with Covid-19 cannot be the same for all subjects, we have to consider a heterogeneous population where individuals have different risks, which leads us to use the risk-based \gt approach introduced in \cite{aprahamian2018}.

Specifically, our work is based on the following assumptions: 
\begin{enumerate}
  \item the tests are imperfect (with classification errors) and can be used on an individual sample as well as on a mixture of samples ;
 \item the population to be screened, which is finite, includes subjects with different known probabilities of being Covid-19 positive, i.e. the prevalence of the disease varies within the population.  
\end{enumerate}

In this paper, we address the \gt design problem (GTDP). The \gtdp consists of determining a partition into groups of a finite population such that the cardinality of each group does not exceed a maximum value and that a testing budget constraint (i.e., representing the required number of tests to screen the population) is satisfied while minimizing a linear combination of the expected number of false negative and false positive classifications. The constraint on maximum group size plays a key role in the context of Covid-19 diagnosis as it avoids the dilution effect, which leads to a decrease in test sensibility. Indeed, the dilution of a single positive sample in a mixture of a large number of negative samples may likely make the positive sample undetectable, which may eventually cause  misclassifications of some subjects. Our main contribution concerns the design, implementation, and evaluation of an exact optimization algorithm to determine such a partition of a population to be tested.

The remainder of the paper is organized as follows.. Section 2 presents a literature review. In Section 3, we describe the \gt design problem and the notation used. The details on the algorithms proposed to tackle the \gt problem are described in Section 4. Computational results are reported in Section 5. Finally, in Section 6, we draw conclusions and discuss future work.

\section{Literature review}

Since the pioneering work of Dorfman \cite{dorfman1943} in \gt in 1943, several relevant scientific contributions were proposed to improve both the \gt procedure and its application to various fields. In the following, we present some relevant contributions in the literature that deal with the \gt design problem.

The well-known Dorfman's \gt procedure is based on two stages: in the first stage, subjects are tested in groups; if a group tests negative, then all subjects in the group are classified as negative; in the second stage, only subjects that belonged to positive groups are re-tested, one-by-one, to single out positives. Dorfman's \gt scheme suppose that the test is perfect, subjects are homogeneous, and the population to be tested is infinite. Several contributions in the literature based on this procedure can be found (see e.g., \cite{sobel1959}, \cite{hwang1975}, \cite{saraniti2006}, \cite{feng2010}, \cite{li2014}). Later some contributions are proposed to extend Dorfman's \gt procedure. Hwang \cite{hwang1975} proposes the earliest works to incorporate subject-specific risk characteristics in \gt design. The extension to the realistic case of imperfect tests is introduced by \cite{blider2012}, \cite{black2012}, \cite{mcmahan2012}, \cite{tebbs2013}, \cite{black2015}. The recent work proposed by Aprahamian et al. \cite{aprahamian2018} relaxed all the unrealistic assumptions introduced initially in Dorfman's \gt procedure. However, they did not take into account the limitation on group size in all their experiments, although this limitation is relevant in some practical situations, as in the case of COVID-19.

\gt has been largely used in several areas other than health, with applications considered in quality control (\cite{sobel1959}), communications (\cite{berger1984}, \cite{wolf1985}), pattern matching (\cite{macula2004}, \cite{clifford2010}), database systems (\cite{cormode2005}), traitor tracing (\cite{meerwald2011}), or machine learning (\cite{zhou2014}). However, in the field of health care, the application of \gt is the most widespread. The relevant applications include blood screening to detect human immunodeficiency virus (HIV), the detection of hepatitis B virus (HBV), and other diseases (\cite{gastwirth1994}, \cite{barlev2010}, \cite{blider2010}, \cite{stramer2011}), the screening of chemical compounds as part of the drug discovery (\cite{zhu2001}), DNA screening (\cite{macula1999}, \cite{du2006}, \cite{cao2016}). 

Recently, some contributions have been proposed on the application of \gt for Covid-19 screening. The prevalence rate of this disease being low with many asymptomatic cases, the application of \gt is of great interest.
Early works in the context of Covid-19 disease focused on determining the ideal group size beyond which the dilution effect causes false negatives. Yelin et al. \cite{yelin2020} show that one can always detect a positive case in the mixed sample of a group of 32 or even 64 subjects under certain conditions. Ben-Ami et al. \cite{benami2020} suggest that the ideal group size should be eight subjects to ensure a reliable group test. Other contributions focused on the application of the \gt scheme but using mainly statistical methods. Curturi et al. \cite{cuturi2020} propose noisy adaptive \gt using Bayesian sequential experimental design with possible application in Covid-19 context. Ghosh et al. \cite{ghosh2020} present Tapestry: a single-round smart pooling technique for Covid-19 testing. The authors claim that Tapestry pooling gives confirmed results in a single round of testing by testing each sample thrice as part of three different pools.

Shental et al. \cite{shental2020} propose a \gt scheme where they pooled $384$ patient samples into $48$ pools, each containing $48$ samples. Each sample was added to six different pools. However, the number of groups and their size are not determined by an optimization algorithm. 

In this paper, we propose a \gt scheme applied to Covid-19 screening using advanced techniques derived from combinatorial optimization. We consider an imperfect test, a finite population and heterogeneous subjects as in \cite{aprahamian2018}. Our main contribution is the addition of a limit on group size to avoid the dilution effect.

\section{Problem definition and notation}

This section relies on the work proposed by Aprahamian et al. \cite{aprahamian2018}.
Let $\mathcal{S} = \{\mathcal S_1,\dots,\mathcal S_N\}$ be the set of subjects to be tested. The risk vector, denoted $\mathbf{p} = (p^1, p^2,\dots,p^N)$,  represents the probability of each individual of being infected by Covid-19. We assume subjects are ordered by increasing values of this risk, i.e. $p^1 \leq p^2 \dots \leq p^N$. 

The \gtdp  consists of finding an {\em optimal feasible} partition of set $\mathcal{S}$. As diagnostic tests are not error-free, we say that a partition is {\em optimal} if it minimizes a convex combination of expected false-negatives and expected false-positives tests. We say that a partition is {\em feasible} if it respects the cardinality constraint for each group, and the budget constraint. We denote by $Se$ the test Sensitivity (i.e., true-positive probability) and by $Sp$ the test Specificity (i.e., true-negative probability). With this notation, we can formulate the \gt design problem as follows:

\begin{mini}[2]
{\Omega\in\mathbf{\Omega}}{ \lambda  \mathbb{E}[FN(\Omega)] + (1-\lambda) \mathbb{E}[FP(\Omega)]}{\label{eq:BM}}{}
\addConstraint{n_i}{\leq L}{ \forall i = 1,\dots, |\Omega|}
\addConstraint{\mathbb{E}[T(\Omega)]}{\leq B},
\end{mini}
where
\begin{itemize}
  \item $\mathbf{\Omega}$ represents the set of all possible partitions of set $\mathcal{S}$;
  \item $\Omega$ represents a single partition of set $\mathcal{S}$, and $|\Omega|$ represents the cardinality of the partition;
  \item $n_i$ denotes the size of group $\Omega_i$ where $\Omega_i\in \Omega$ states that $\Omega_i$ is a set belonging to partition $\Omega$;
  \item $\mathbb{E}[FN(\Omega)]$ is the expected number of false-negative classifications;
  \item $\mathbb{E}[FP(\Omega)]$ corresponds to the expected number of false-positive classifications;
  \item $\mathbb{E}[T(\Omega)]$ represents the expected number of tests to perform;
  \item $\lambda \in [0,1]$ is a weight, set by the decision maker, that controls the importance of false negatives versus false positives in the objective function;
  \item $L$ is the maximum group cardinality;
  \item $B$ is the test budget, i.e., the maximal number of tests that can be performed.
\end{itemize}

The objective function in model \req{eq:BM} minimizes the weighted sum of the expected number of false-negative and false-positive classifications. The constraints state that the size of each group cannot exceed the maximum group size and that the expected number of tests must be smaller or equal than a given testing budget.

\subsection{Expected number of false negatives, positives and tests}

Let $m \in \{1,\dots,|\mathcal S|\}$. We denote by $\mathcal S_m$ the subject corresponding to position $m$ in $\mathcal S$. Let $I^m$ denote the indicator random variable corresponding to the true-positive status of subject $\mathcal S_m \in \mathcal{S}$. For a partition $\Omega\in\mathbf{\Omega}$, let $FN^m(\Omega)$ and $FP^m(\Omega)$, denote the indicator random variables, respectively corresponding to the false-negative classification and false-positive classification of subject $\mathcal S_m$. Similarly, let $N_i^+(\Omega_i)$, $FN_i(\Omega_i)$ and $FP_i(\Omega_i)$ respectively denote the counterparts of these random variables for group $\Omega_i\in\Omega$, $\forall i = 1,\dots, |\Omega|$. Note that, for a partition $\Omega$, $\Omega^I$ and $\Omega^G$ respectively correspond to the sets of subjects to be tested individually and in groups. The computation of the expected number of false negatives, positives and tests is presented in Sections~\ref{sec::subfn}--\ref{sec::subtests}.


\subsubsection{Expected number of false negatives}\label{sec::subfn}

In individual testing, a truly positive subject is falsely classified as negative if the test outcome is negative. In contrast, in \gt, a truly positive subject is falsely classified as negative if (i) the group test outcome is negative, or (ii) the group test outcome is positive, and the subject's subsequent individual test outcome is negative. Then, given a partition $\Omega$, for any subject $\mathcal S_m  \in \mathcal{S}$ with risk $p^m$, we have: \\ \\
\bigskip
$
\mathbb{E}[FN^m]=\mathbb{E}[FN^m|I^m = 1]P(I^m = 1) + \mathbb{E}[FN^m|I^m = 0]P(I^m = 0)\\
$

\noindent \hspace*{1.5cm}$
=\left\{
\begin{array}{ll}
(1-Se)p^m+0, & \mbox{if} \quad \mathcal S_m\in \Omega^I, \\
(Se(1-Se)+(1-Se))p^m+0, & \mbox{if}\quad  \mathcal S_m \in \Omega^G,\\
\end{array}
\right.
$\\\\

\noindent leading to:
\noindent$
\mathbb{E}[FN^m]=\left\{
\begin{array}{ll}
(1-Se)p^m, & \mbox{if} \quad \mathcal S_m\in \Omega^I, \\
(1-Se^2)p^m, & \mbox{if}\quad  \mathcal S_m\in \Omega^G.
\end{array}
\right.
$\\

\noindent Then, the expected number of false-negative classifications for group $i$ is given by:\\

\noindent$
\mathbb{E}[FN_i(\Omega_i)]=\left\{
\begin{array}{ll}
(1-Se)\sum_{m \in \Omega_i}p^m, & \mbox{if} \quad n_i=1, \\
(1-Se^2)\sum_{m \in \Omega_i}p^m, & \mbox{otherwise,}
\end{array}
\right.
$\\

\noindent and the expected number of false-negative classifications for all subjects in set $\mathcal{S}$ is given by:
%
\begin{align}
\mathbb{E}[FN(\Omega)] = (1-Se) \sum_{m \in \Omega^I}^{} p^m + (1-Se^2) \sum_{m \in \Omega^G}^{} p^m.
\end{align} 

\subsubsection{Expected number of false positives}

In individual testing, a truly negative subject is falsely classified as positive if the test outcome is positive, whereas in \gt, a truly negative subject is falsely classified as positive if the group test outcome is positive and the subject's subsequent individual test outcome is positive. Then, given a partition $\Omega$, for any individually tested subject $\mathcal S_m \in \Omega^I$, we can write: \\ 

\noindent$
\mathbb{E}[FP^m]=\mathbb{E}[FP^m|I^m = 1]P(I^m = 1) + \mathbb{E}[FP^m|I^m = 0]P(I^m = 0)\\
$

\noindent \hspace*{1.5cm}$
= 0 + (1-Sp)(1-p^m),\\
$

\noindent and for any subject $\mathcal S_m \in \Omega^G$ grouped in some set $\Omega_{i}$ such that $i\in \{1,\dots,|\Omega|\}$, $n_i > 1$ (i.e., $\mathcal S_m\in \Omega_i$), we have: \\

\noindent$
\mathbb{E}[FP^m]=\mathbb{E}[FP^m|I^m = 1]P(I^m = 1) + \mathbb{E}[FP^m|I^m = 0]P(I^m = 0)\\
$

\noindent \hspace*{1.5cm}$
= 0 + \bigg[(1-Sp)^2 \prod_{k\in \Omega_i\setminus\{m\}}(1-p^k) + Se(1-Sp)\bigg(1-\prod_{k\in \Omega_i\setminus\{m\}}(1-p^k)\bigg)\bigg](1-p^m)\\
$

\noindent \hspace*{1.5cm}$
= (1-Sp) \bigg[Se-(Se+Sp-1) \prod_{k\in \Omega_i\setminus\{m\}}(1-p^k)\bigg](1-p^m)\\
$

\noindent \hspace*{1.5cm}$
= (1-Sp)Se(1-p^m)-(1-Sp)(Se+Sp-1) \prod_{k\in \Omega_i}(1-p^k),\\
$

 \noindent leading to: \\
 
\noindent $
\mathbb{E}[FP_i(\Omega_i)] = \left\{
\begin{array}{ll}
(1-Sp)(1-p^m), & \mbox{if } \quad \mathcal S_m \in \Omega^I, \\
(1-Sp)Se(1-p^m)-(1 - Sp).(Se+Sp-1) \prod_{k\in \Omega_i}(1 - p^k), & \mbox{if} \quad \mathcal S_m \in \Omega^G.
\end{array}
\right.
$\\

 \noindent Then, the expected number of false-positive classifications for group $\Omega_i$ is given by: 
\begin{align}
\mathbb{E}[FP_i(\Omega_i)] = \left\{
\begin{array}{ll}
(1-Sp) \sum_{m\in \Omega_i}^{} (1-p^m), & \mbox{if } n_i=1, \\
(1-S_p)Se \sum_{m\in \Omega_i}^{} (1-p^m) - ni(1 - Sp). (Se + Sp - 1) \prod_{m\in \Omega_i}^{} (1 - p^m), & \mbox{otherwise,}
\end{array}
\right.
\end{align}

\noindent and the expected number of false-positive classifications for all subjects in set $\mathcal{S}$ is given by $\mathbb{E}[FP(\Omega)] = \sum_{i}^{} \mathbb{E}[FP_i(\Omega_i)]$.

\subsubsection{Expected number of tests}\label{sec::subtests}

In individual testing, the number of tests per subject is always one. In \gt, the number of tests depends on the outcome of the group test: if the group test outcome is negative, then only one test is performed for the entire group, and if the group test outcome is positive, an additional individual test is performed for each subject in the group. Given a partition $\Omega$, the expected number of tests for group $\Omega_i$, $i=\{1,\dots,|\Omega|\}$, is $1$ if $n_i = 1$ (i.e., individual testing), and if $n_i > 1$, we can write: \\

\noindent $\mathbb{E}[T_i(\Omega_i)] = \sum_{k=0}^{ni} \mathbb{E}[T_i(\Omega_i)|N_i^+(\Omega_i)=k]P(N_i^+(\Omega_i)=k)\\
$

\noindent \hspace*{1.5cm} $ = \mathbb{E}[T_i(\Omega_i)|N_i^+(\Omega_i)=0]P(N_i^+(\Omega_i)=0)\\
\hspace*{2cm}+\sum_{k=1}^{ni} \mathbb{E}[T_i(\Omega_i)|N_i^+(\Omega_i)=k]P(N_i^+(\Omega_i)=k)\\
$

\noindent \hspace*{1.5cm} $ = (Sp+(1-Sp)(1+n_i))P(N_i^+(\Omega_i)=0)\\
\hspace*{2cm}+\sum_{k=1}^{ni}(1-Se+Se(1+n_i))P(N_i^+(\Omega_i)=k)\\
$

\noindent \hspace*{1.5cm} $ = 1+n_i\bigg(Se-(Se+Sp-1)\prod_{m\in \Omega_i}(1-p^m)\bigg).\\
$

\noindent Thus, 
\begin{align}
\mathbb{E}[T_i(\Omega_i)] = \left\{
\begin{array}{ll}
1, & \mbox{if } n_i=1, \\
1+n_i\bigg(Se - (Se + Sp - 1) \prod_{m\in \Omega_i}^{} (1 - p^m)\bigg), & \mbox{otherwise,}
\end{array}
\right.
\end{align}
and the expected number of tests needed for all subjects in set $\mathcal{S}$ is given by:
 
\noindent $\mathbb{E}[T(\Omega)] = \sum_i \mathbb{E}[T_i(\Omega_i)]$.

For more details on how the expressions of the expected number of false negatives, positives and tests are computed, we refer the interested readers to \cite{aprahamian2018}.

\section{An exact algorithm for the \gtdp}\label{sec:algo}


Before presenting the algorithm that we develop to partition the sets of individuals in groups, let us introduce the definition of an {\em ordered partition} (\cite{aprahamian2018}).

A partition, $\Omega = (\Omega_i)_{i=1,\dots,|\Omega|}$, is said to be an {\em ordered partition} if it follows the ordered set $\mathcal S = \{\mathcal S_1,\dots,\mathcal S_N\}$, that is, $\Omega_1 = \{\mathcal S_1,\dots,\mathcal S_{n_1}\}$, $\Omega_2 = \{\mathcal S_{n_1+1},\dots,\mathcal S_{n_1+n_2}\}$,\dots, $\Omega_{|\Omega|} = \{\mathcal S_{\sum_{i=1}^{|\Omega|-1} n_i  + 1},\dots,\mathcal S_N\}$, where $|\Omega|\in \{1,\dots,N\}$ and $n_i \in \mathbb{Z}^+$, $i = 1,\dots,|\Omega|$.

Aprahamian et al.~\cite{aprahamian2018} showed that
the following properties hold for model \req{eq:BM} when only the maximum budget constraint is considered:
\begin{enumerate}
    \item  There exists an optimal partition that is an {\em ordered partition} of $\mathcal S$.
    \item  If subject $\mathcal S_m$ is individually tested in an optimal partition, then it is optimal to individually test all subjects having a risk higher than $p^m$.
\end{enumerate}

These properties remain valid when we consider the maximum group size constraints. Indeed, the proofs of both properties are based on the same type of argument. Let us consider the first property. Suppose that there exists an optimal partition $\Omega^*$ that does not follow an ordered partition of $\mathcal S$. Two groups of $\Omega^*$ violating the ordered partition property are identified. A new partition $\hat{\Omega}$ is then constructed by interchanging subjects between the two groups to satisfy the ordered partition property. All other groups of $\Omega^*$ remain unchanged in $\hat{\Omega}$. The key observation is that the subgroups of subjects interchanged between the two groups have the same cardinality in the proof by Aprahamian et al.~\cite{aprahamian2018}. This means that if $\Omega^*$ satisfies the cardinality constraints on the groups, $\hat{\Omega}$ also satisfies them. Therefore, the rest of the proof by Aprahamian et al.~\cite{aprahamian2018} is valid. The proof of the second property follows the same lines and is therefore valid when we impose the cardinality constraints on the size of the group.

Since the properties given above still hold, problem \req{eq:BM} reduces to solve a Constrained Shortest Path Problem \cite{garcia2009} over the directed graph $G = (V,A)$ defined as follows:

\begin{itemize}
\item the node set $V = \mathcal S \cup \{\mathcal S_{N+1}\}$, contains one node per subject in $\mathcal S$, and an artificial node $\mathcal S_{N+1}$ that represents a subject with risk $1$;
\item the arc set $A = \{(\mathcal S_i,\mathcal S_j) : \mathcal S_i, \mathcal S_j \in V, i < j, \mathbb{E}[T_i(\Omega_{i-j})] \leq B, j - i \leq L\}$, contains arcs $(\mathcal S_i,\mathcal S_j)$ that represent the corresponding groups $\Omega_{i-j} = \{\mathcal S_i, \mathcal S_{i+1}, \dots, \mathcal S_{j-1}\}$ that are feasible with respect to the maximum group size constraint and to the budget constraint; 
\item the cost of the arc $(\mathcal S_i,\mathcal S_j)\in A$ is equal to $C_{ij} =  \lambda  \mathbb{E}[FN_i(\Omega_{i-j})] + (1-\lambda) \mathbb{E}[FP_i(\Omega{}_{i-j})]$.
\end{itemize}

Aprahamian et al. \cite{aprahamian2018} showed that each path from node $S_1$ to node $S_{N+1}$ corresponds to an ordered partition of set $\mathcal S$. 
To solve \req{eq:BM} as a Constrained Shorted Path Problem, we propose an algorithm derived from the exact algorithm of Feillet et al. \cite{feillet2004}, which is a dynamic programming based procedure based on a label correcting strategy. Due to the fact that the cost of the arcs of the graph $G = (V, A)$ are all positive, we do not have to impose elementarity reducing the complexity of the algorithm proposed in Feillet et al. \cite{feillet2004}. Moreover, the acyclic structure of the graph helps in reducing the computational burden to obtain the optimal partition $\Omega$ of $\mathcal S$.

\subsubsection*{Paths and labels}

Each feasible path between the sink ($\mathcal S_1$) and a node ($\mathcal S_i$) corresponds to a label $l^i$.
Such a label is a triplet $l = (pred,C, R)$ containing a pointer to the label of its predecessor ($pred$) with respect to the order of subjects in the list $\mathcal S$, the cost of the path ($C$), and the consumption of the resource ($R$), which corresponds to the expected number of tests. We do not need to introduce an additional resource related to the size of the groups since groups that exceed the maximum size are excluded when constructing the graph.

Let $l_1^i$ and $l_2^i$ be two distinct paths from the sink ($\mathcal S_1$) to a node ($\mathcal S_i$). We say that $l_1^i < l_2^i$ (label $l_1^i$ lower than $l_2^i$) if  $C^{l_1^i}  \leq C^{l_2^i}$.
We store for every node ($\mathcal S_i$) in the acyclic graph the list of labels  $\Lambda_i = \{l_1^i, l_2^i, .., l_k^i$\} corresponding to all the possible paths from ($\mathcal S_1$) to ($\mathcal S_i$), with $l_1^i$ $<$ $l_2^i$ $<$ \dots $<$ $l_k^i.$

Let $\mathcal S_j$ ($j > 1$) be a node. Let $\mathcal S' = \{\mathcal S_i \in \mathcal S : i \in \{1,\dots,j-1\}\}$ be the list of predecessors of $\mathcal S_j$. Let $\Lambda_i$ be the list of labels on every predecessor $\mathcal S_i \in \mathcal S'$. Each label $l^j$, corresponding to a feasible path from $\mathcal S_1$ to $\mathcal S_j$, is computed as follows :

For all label $l^i \in \Lambda_i$ :  $l^j = (l^i, C^{l^j}, R^{l^j})$ such as $C^{l^j} = C^{l^i} + C_{ij}$, $R^{l^j} = R^{l^i} + R_{ij}$ if $R^{l^j} + R_{ij} \leq B$ with $R_{ij}= \mathbb{E}[T_i(\Omega_{i-j})]$. For $j = 1$, there is no predecessors. Thus, $\Lambda_1$ contains only the label $(NIL, 0,0)$.

As the number of labels on every node can increase exponentially, we only keep in memory the non dominated paths thanks to the following dominance rule. 

\subsubsection*{Dominance rule}

Let $l_1^i$ and $l_2^i$ be two distinct paths from the sink ($\mathcal S_1$) to a node ($\mathcal S_i$). $l_1^i$ dominates $l_2^i$ if and only if $l_1^i \leq l_2^i$ and $R^{l_1^i} \leq R^{l_2^i}$.
This dominance rule allows us to save only Pareto-optimal paths all along the execution of the algorithm.




\subsubsection*{Algorithm}

Algorithm \textbf{\textit{GTCSPP}} (\gt Constrained Shortest Path Problem) is described in Algorithm~\ref{gtcspp}. It applies the label correcting algorithm by traversing nodes in topological order on the implicit acyclic graph $G$. Thus, when processing a node $S_j$, all the nodes $S_i \in \{S : i < j\}$ that precede it in topological order have their final labels. Labels of this node are, therefore, only computed once.

Starting by processing node $S_1$, at each step of the algorithm, the first unprocessed node in the topological order imposed by $\mathcal S$ is selected. Then, all its labels are extended to each of its successors thanks to Algorithm \ref{extend} (\textbf{\textit{ExtendAndDominance}}). Its outgoing arcs define the set of successors of a node.

Before inserting a label in the list of labels associated with the corresponding node, we check if the label is dominated by an already present label. In this case, the new label is discarded. If the new label is not dominated, we check if it dominates already present labels, and in the affirmative case, we discard the dominated labels.

To obtain the optimal partition after computing the shortest path with resource constraint between the first node and the artificial node, we browse the resulting path backward, as described in Algorithm~\ref{construct}.

\begin{algorithm}[H]
\caption{GTCSPP}
\label{gtcspp}
\begin{algorithmic}[1]
\Require
\\
 \begin{itemize}
    \item[-] $\mathcal S$: ordered list of subjects of size $N$,
    \item[-] $B$ : number of available tests.
\end{itemize}
\Ensure partition $P$ of the subjects $\mathcal S$
\Function{GTCSPP}{$\mathcal S, B, M$} 
\State  Add an artificial subject at the end of S with $risk = 1$;
\State $\Lambda_1 \gets (NIL, 0,0)$;
\For {$j=2,\dots,N+1$}
\State $\Lambda_j \gets \emptyset$;
\State $i \gets 1$;
\While {{($i <j$)} {and $((\mathcal S_i,\mathcal S_j) \in A$)}}
\State  $R_{ij} \gets \mathbb{E}[T_i(\Omega_{i-j})]$;
\State $C_{ij} \gets \lambda  \mathbb{E}[FN(\Omega_{i-j})] + (1-\lambda) \mathbb{E}[FP(\Omega_{i-j})]$;
\For {$l \in \Lambda_i$}
\If{$R^l + R_{ij} \leq B$}
\State $l'\gets (l, C^l + C_{ij}, R^l + R_{ij})$;
\State $\Lambda_j \gets ExtendAndDominance(\Lambda_j, l)$; 
\EndIf 
\EndFor
\State $i \gets i + 1$;
\EndWhile
\EndFor
\If{$\Lambda_{N+1} \neq \emptyset$} 
\State $l \gets \Lambda_{N+1}^1$; 
\State $I \gets \emptyset$;
\While {$pred^l \neq $NIL}
\State $I \gets I \cup \{pred^l\}$;
\State $l \gets pred^l$;
\EndWhile
\EndIf
\State Remove the artificial subject at the end of $\mathcal S$;
\State $P \gets RecoverPartition(\mathcal S, I)$;
\State \textbf{return} $P$;
\EndFunction
\end{algorithmic}
\end{algorithm}

\begin{algorithm}[H]
\caption{Extend and dominance}
\label{extend}
\begin{algorithmic}[1]
\Require \\
\begin{itemize}
    \item[-] $\Lambda_j$: sorted list of size $K$ of labels at a node $j$ 
    \item[-]  $l$ : the new label to be added
    \item[-]  dominates($l_1, l_2$) : returns $true$ if $R^{l_1} <= R^{l_2}$ when $l_1 < l_2$
\end{itemize} 
\Ensure insert $l$ in the right position in $\Lambda_j$ if not dominated and remove dominated labels if any
\Function{ExtendAndDominance}{$\Lambda_j, l$} 
\State $index \gets 0$;
\If{$\Lambda_j = \emptyset$}
    \State $index \gets 0$;
\Else
   \State $i\gets 1$;
    \While{$i \leq K$}
        \If{$l < \Lambda_j^i$}
            \State $index \gets i$;
            \If{$dominates(l, \Lambda_j^i$)}
                \While{$i \leq K$};
                    \If{$dominates(l, \Lambda_j^i)$}
                    \State remove $\Lambda_j^i$;
                    \Else
                    \State $i \gets i + 1$;
                    \EndIf
                \EndWhile
            \EndIf
            \State break;
        \Else
            \If{$dominates(\Lambda_j^i, l)$}
                \State $index \gets 0$;
                \State break;
            \Else
                \State $index \gets i$;
                \State $i \gets i + 1$;
            \EndIf
        \EndIf
    \EndWhile
\EndIf
\If{$index \neq 0$}
    \State insert $l$ at $index$ in $\Lambda_j$
\EndIf
\State \textbf{return} $\Lambda_j$;
\EndFunction
\end{algorithmic}
\end{algorithm}

\begin{algorithm}[H]
\caption{Recover partition}
\label{construct}
\begin{algorithmic}[1]
\Require
\\
\begin{itemize}
    \item[-] $\mathcal S$ : list of subjects
    \item[-] $I$ : list of size $M$ of labels corresponding to first elements in every group
\end{itemize}
\Ensure The corresponding partition $P$
\Function{RecoverPartition}{$\mathcal S, I$}
\For {$i=M,\dots,1$} 
    \State $P \gets P \cup \emptyset$;
    \For {$j=I^i,\dots,I^{i-1}$} 
        \State $P^{M-i} \gets P^{M-i}\cup \mathcal S_j$;
    \EndFor
\EndFor
\State \textbf{return} $P$;
\EndFunction
\end{algorithmic}
\end{algorithm}

\begin{figure}[H]
\caption{An example of graph with 8 subjects, $\mathcal S_9$ is the artificial one.}
\label{graph}
  \footnotesize
  \centering
  \begin{tikzpicture}[shape=circle,auto]
    \node[draw] (S1) at (0,0){$\mathcal S_1$};
    \node[draw] (S2) at (2,0) {$\mathcal S_2$};
    \node[draw] (S3) at (4,0) {$\mathcal S_3$};
    \node[draw] (S4) at (6,0){$\mathcal S_4$};
    \node[draw] (S5) at (8,0) {$\mathcal S_5$};
    \node[draw] (S6) at (10,0) {$\mathcal S_6$};
    \node[draw] (S7) at (12,0) {$\mathcal S_7$};
    \node[draw] (S8) at (14,0) {$\mathcal S_8$};
    \node[draw] (S9) at (16,0) {$\mathcal S_9$};

    \path[->] (S1) edge (S2);
    \path[->] (S2) edge (S3);
    \path[->] (S3) edge (S4);
    \path[->] (S4) edge (S5);
    \path[->] (S5) edge (S6);
    \path[->] (S6) edge (S7);
    \path[->] (S7) edge (S8);
    \path[->] (S8) edge (S9);
 
    \path[red, thick,bend left,->] (S1) edge (S6);
    \path[red, thick,bend left,->] (S6) edge (S9);
     \path[thick,bend right,->] (S1) edge (S5);
    \path[thick,bend right,->] (S5) edge (S7);
    \path[thick,bend right,->] (S7) edge (S9);

  \end{tikzpicture}
\end{figure}

Figure~\ref{graph} shows the graph $G=(V,A)$ corresponding to an example of 8 subjects $\mathcal S = \{\mathcal S_1, \mathcal S_2,\dots, \mathcal S_8\}$. Let $ \mathbf{p}=\{0.01, 0.02, 0.03, 0.04, \allowbreak 0.05, 0.06, 0.07, 0.08\}$ be the corresponding risk vector.

The costs $C_{ij}$ of the arcs $(\mathcal S_i, \mathcal S_j) \in A, i < j$, are computed as explained above. Let $Se = Sp = 0.75$, $\lambda = 0.6$ and $B = 6$ . The optimal solution computed by Algorithm~\ref{gtcspp} is $\Omega^* = \{\{\mathcal S_1,\mathcal S_2,\mathcal S_3,\mathcal S_4,\mathcal S_5\}, \{\mathcal S_6,\mathcal S_7, \mathcal S_8\}\}$, with an expected cost $C =0.311$ and an expected number of tests $R = 5.444$. The other depicted path corresponds to the feasible partition $\Omega = \{\{\mathcal S_1,\mathcal S_2,\mathcal S_3,\mathcal S_4\},\{\mathcal S_5,\mathcal S_6\},\allowbreak \{\mathcal S_7, \mathcal S_8\}\}$ with with an expected cost $C =0.332$ and an expected number of tests $R = 4.647$.

\section{Computational experiments}

In this section, we evaluate the  \textbf{\textit{GTCSPP}} algorithm proposed to tackle the \gtdp in the case of Covid-19 screening. 
We ran our algorithm on instances generated from the data-set provided by the French national health care agency (Sant\'e publique France). We used data related to Covid-19 screening tests carried out in city laboratories.
These data concern all 101 French departments. They cover 11 weeks, from week 11 to week 21 of 2020 (March 10 to May 24). This period corresponds to the peak of the first epidemic in France.

The data set contains the number of tests performed, \textbf{404,550}, the number of people tested, \textbf{404,046} and the number of people tested positive, \textbf{39,830}. These data are classified by gender (women and men) and by age group (five age groups). Table~\ref{femtable} and~\ref{maltable} report the numbers of women and men tested, the number of women and men tested positive, and the risk of Covid-19 for each subgroup.

\begin{table}
\centering
\scalebox{0.8}{
 \begin{threeparttable}
\caption{The number of women tested, the number of women tested positive and the risk of Covid-19 for each age group of women}\label{femtable}
\begin{tabular}{|l| c| c c c c c c c c c c c|}
\hline
\backslashbox{Age group}{Week} &  & 11 & 12 & 13 & 14 & 15 & 16 & 17 & 18 & 19 & 20 & 21\\ 
\hline
\multirow{3}{*}{Women under 15 years}  & pos & 0 &  1 & 10 & 16 & 12 & 4 & 5 & 2 & 3 & 12 & 5\\
						& tot & 22 & 86 & 144 & 211 & 248 & 226 & 245 & 213 & 411 & 1004 & 1014\\
						& risk & 0 & 1.16 & 6.94 & 7.58 & 4.84 & 1.77 & 2.04 & 0.94 & 0.73 & 1.2 & 0.49\\
\hline
\multirow{3}{*}{Women between 15-44 years} & pos & 44 & 573 & 1856 & 2047 & 1575 & 859 & 571 & 257 & 188 & 148 & 79\\
						& tot &  272 & 2798 & 8532 & 11958 & 12026 & 11788 & 11752 & 7951 & 9155 & 11355 & 8708\\
						& risk & 16.18 & 20.48 & 21.75 & 17.12 & 13.1 & 7.29 & 4.86 & 3.23 & 2.05 & 1.3 & 0.91\\
\hline
\multirow{3}{*}{Women between 45-64 years} & pos & 43 & 525 & 1634 & 1896 & 1565 & 833 & 477 & 231 & 139 & 112 & 47\\
						& tot & 201 & 1841 & 6289 & 9395 & 10281 & 9938 & 9806 & 6798 & 7851 & 9582 & 7298\\
						& risk & 21.39 & 28.52 & 25.98 & 20.18 & 15.22 & 8.38 & 4.86 & 3.4 & 1.77 & 1.17 & 0.64\\
\hline
\multirow{3}{*}{Women between 65-74 years} & pos & 11 & 117 & 366 & 390 & 351 & 221 & 117 & 42 & 48 & 32 & 24\\
						& tot & 37 & 328 & 1209 & 1753 & 2164 & 2089 & 2140 & 1440 & 1842 & 2801 & 2365\\
						& risk & 29.73 & 35.67 & 30.27 & 22.25 & 16.22 & 10.58 & 5.47 & 2.92 & 2.61 & 1.14 & 1.01\\
\hline
\multirow{3}{*}{Women 75 years and over} & pos & 19 & 237 & 782 & 1307 & 2383 & 1656 & 965 & 337 & 281 & 268 & 141\\
						& tot & 71 & 824 & 2762 & 3917 & 9250 & 12581 & 12418 & 7928 & 7837 & 7266 & 4631\\
						& risk & 26.76 & 28.76 & 28.31 & 33.37 & 25.76 & 13.16 & 7.77 & 4.25 & 3.59 & 3.69 & 3.04\\
\hline
\end{tabular}
   \begin{tablenotes}
        \item pos : number of positive women.
        \item tot : total number of women tested.
        \item risk : Covid-19 prevalence rate ($\%$).
    \end{tablenotes}
\end{threeparttable}
}
\end{table}

\begin{table}
\centering
\scalebox{0.8}{
 \begin{threeparttable}
\caption{The number of men tested, the number of men tested positive and the risk of Covid-19 for each  age group of men}\label{maltable}
\begin{tabular}{|l| c| c c c c c c c c c c c|}
\hline
\backslashbox{Age group}{Week} &  & 11 & 12 & 13 & 14 & 15 & 16 & 17 & 18 & 19 & 20 & 21\\ 
\hline
\multirow{3}{*}{Men under 15 years}  & pos & 3 &  10 & 25 & 41 & 21 & 13 & 18 & 11 & 8 & 21 & 9\\
						& tot & 24 & 103 & 171 & 208 & 204 & 246 & 268 & 216 & 387 & 1160 & 1161\\
						& risk & 12.5 & 9.71 & 14.62 & 19.71 & 10.29 & 5.28 & 6.72 & 5.09 & 2.07 & 1.81 & 0.78\\
\hline
\multirow{3}{*}{Men between 15-44 years} & pos & 38 & 291 & 837 & 804 & 573 & 362 & 234 & 140 & 108 & 80 & 55\\
						& tot &  170 & 1066 & 3190 & 4655 & 4926 & 4552 & 4700 & 3543 & 4529 & 7449 & 5921\\
						& risk & 22.35 & 27.3 & 26.24 & 17.27 & 11.63 & 7.95 & 4.98 & 3.95 & 2.38 & 1.07 & 0.93\\
\hline
\multirow{3}{*}{Men between 45-64 years} & pos & 32 & 399 & 1213 & 1119 & 760 & 416 & 265 & 124 & 98 & 85 & 45\\
						& tot & 119 & 1089 & 3530 & 4792 & 4846 & 4285 & 4599 & 3394 & 4200 & 6172 & 4933\\
					  & risk & 26.89 & 36.64 & 34.36 & 23.35 & 15.68 & 9.71 & 5.76 & 3.65 & 2.33 & 1.38 & 0.91\\
\hline
\multirow{3}{*}{Men between 65-74 years} & pos & 18 & 200 & 551 & 477 & 342 & 205 & 126 & 43 & 47 & 22 & 20\\
						& tot & 64 & 468 & 1501 & 1851 & 1918 & 1861 & 2030 & 1430 & 1713 & 2666 & 2259\\
						& risk & 28.13 & 42.74 & 36.71 & 25.77 & 17.83 & 11.02 & 6.21 & 3.01 & 2.74 & 0.83 & 0.89\\
\hline
\multirow{3}{*}{Men 75 years and over} & pos & 13 & 192 & 654 & 784 & 874 & 526 & 310 & 133 & 93 & 87 & 55\\
						& tot & 54 & 625 & 1746 & 2282 & 3536 & 4225 & 4364 & 2960 & 2919 & 3291 & 2423\\
						& risk & 24.07 & 30.72 & 37.46 & 34.36 & 24.72 & 12.45 & 7.1 & 4.49 & 3.19 & 2.64 & 2.27\\
\hline
\end{tabular}
   \begin{tablenotes}
        \item pos : number of positive men.
        \item tot : total number of men tested.
        \item risk : Covid-19 prevalence rate ($\%$).
    \end{tablenotes}
\end{threeparttable}
}
\end{table}

\subsection{Instance generation}

An instance of the \gtdp is composed of a set of subjects that must be tested. Each subject has a risk, i.e., Covid-19 prevalence rate. The prevalence of a disease corresponds to the number of cases in a population at a given point in time, including both new and old cases.
The risk of each subject is determined according to his/her gender age group. We compute each gender age group risk weekly (see Table~\ref{femtable} and \ref{maltable}) as Covid-19 prevalence vary along time. Note that the French national health care agency initially defined the partition of the data set into age groups by gender. To ensure that it is meaningful to use different risks by gender and age group, we need to verify whether the risks are statistically different by performing  $\chi^2$ tests on the data for the entire period of the epidemic. Results of statistical tests are reported in Tables~\ref{femtest}, \ref{homtest} and \ref{femhomtest}. These tables report the $\chi^2$ test results on the significance of women, men, and women and men Covid-19 prevalences. Columns correspond to weeks of Covid-19 screening tests and rows report the comparisons between a gender age group prevalence and the other ones. A green cell in the table indicates that the prevalence of the gender-age group this week is significantly different from the prevalences of all other gender-age groups. A red cell indicates that the prevalence of the gender-age group is not significantly different from at least one of the other gender-age group prevalences.
The results in Tables~\ref{femtest}, \ref{homtest} and \ref{femhomtest} show that the prevalences of the $10$ gender-age groups are not statistically different during the epidemic period in France. Therefore, we group the same age group of different genders, reducing the numbers of groups from $10$ to $5$. The partition of the number of people tested, the number of people tested positive and the risk of Covid-19 for each of the $5$ age groups is reported in Table~\ref{agetable}.

\begin{table}
\centering
\scalebox{0.8}{
 \begin{threeparttable}
\caption{$\chi^2$ significance test of prevalence between different age groups of women}
\label{femtest}
\begin{tabular}{|l| c| c |c |c| c| c |c |c |c |c |c|}
\hline
\backslashbox{Age group}{Week}  & 11 & 12 & 13 & 14 & 15 & 16 & 17 & 18 & 19 & 20 & 21\\ 
\hline
W-15 vs 4 others  &  \cellcolor{green} &  \cellcolor{green} & \cellcolor{green} & \cellcolor{green} & \cellcolor{green} & \cellcolor{green} & \cellcolor{green} & \cellcolor{green} & \cellcolor{green} & \cellcolor{red} & \cellcolor{red}\\
\hline
W15-44 vs 3 others &  \cellcolor{red} &  \cellcolor{green} & \cellcolor{green} & \cellcolor{green} & \cellcolor{green} & \cellcolor{green} & \cellcolor{red} & \cellcolor{red} & \cellcolor{red} & \cellcolor{red} & \cellcolor{red}\\
\hline
W45-64 vs 2 others &  \cellcolor{red} &  \cellcolor{red} & \cellcolor{green} & \cellcolor{red} & \cellcolor{red} & \cellcolor{green} & \cellcolor{red} & \cellcolor{red} & \cellcolor{green} & \cellcolor{red} & \cellcolor{red}\\ 
\hline
W65-74 vs W75+  & \cellcolor{red} &  \cellcolor{green} & \cellcolor{red} & \cellcolor{green} & \cellcolor{green} & \cellcolor{green} & \cellcolor{green} & \cellcolor{green} & \cellcolor{green} & \cellcolor{green} & \cellcolor{green}\\
\hline
\end{tabular}
   \begin{tablenotes}
        \item W-15, W15-44, W45-64, W65-74, W75+ : Women under 15, between 15-44, 45-64, 65-74 and 75 years and over.
        \item 4 others : W15-44, W45-64, W65-74, W75+.
        \item 3 others : W45-64, W65-74, W75+.
        \item 2 others : W65-74, W75+.
        \item \colorbox{green}{ green week 11}: means W-15 has a prevalence of Covid-19 significantly different to the ones of all other age groups for the week 11 (see the first row).
        \item \colorbox{red}{ red week 20}: means W-15 has a prevalence of Covid-19 that is not significantly different to at least one of the others for the week 20 (see the first row).
    \end{tablenotes}
\end{threeparttable}
}
\end{table}

\begin{table}
\centering
\scalebox{0.8}{
 \begin{threeparttable}
\caption{$\chi^2$ significance test of prevalence between different age groups of men}
\label{homtest}
\begin{tabular}{|l| c| c |c |c| c| c |c |c |c |c |c|}
\hline
\backslashbox{Age group}{Week}  & 11 & 12 & 13 & 14 & 15 & 16 & 17 & 18 & 19 & 20 & 21\\ 
\hline
M-15 vs 4 others  &  \cellcolor{red} &  \cellcolor{green} & \cellcolor{green} & \cellcolor{red} & \cellcolor{red} & \cellcolor{red} & \cellcolor{red} & \cellcolor{red} & \cellcolor{red} & \cellcolor{red} & \cellcolor{red}\\
\hline
M15-44 vs 3 others &  \cellcolor{red} &  \cellcolor{red} & \cellcolor{green} & \cellcolor{green} & \cellcolor{green} & \cellcolor{green} & \cellcolor{red} & \cellcolor{red} & \cellcolor{red} & \cellcolor{red} & \cellcolor{red}\\
\hline
M45-64 vs 2 others &  \cellcolor{red} &  \cellcolor{green} & \cellcolor{red} & \cellcolor{green} & \cellcolor{green} & \cellcolor{red} & \cellcolor{red} & \cellcolor{red} & \cellcolor{red} & \cellcolor{green} & \cellcolor{red}\\ 
\hline
M65-74 vs M75+  & \cellcolor{red} &  \cellcolor{green} & \cellcolor{red} & \cellcolor{green} & \cellcolor{green} & \cellcolor{red} & \cellcolor{red} & \cellcolor{green} & \cellcolor{red} & \cellcolor{green} & \cellcolor{green}\\
\hline
\end{tabular}
   \begin{tablenotes}
        \item M-15, M15-44, M45-64, M65-74, M75+ : Male under 15, between 15-44, 45-64, 65-74 and 75 years and over.
        \item 4 others : M15-44, M45-64, M65-74, M75+.
        \item 3 others : M45-64, M65-74, M75+.
        \item 2 others : M65-74, M75+.
        \item \colorbox{green}{ green week 18}: means M65-74 has a prevalence of Covid-19 significantly different to the one of M75+ for the week 18 (see the last row).
         \item \colorbox{red}{ red week 11}: means M-15 has a prevalence of Covid-19 that is not significantly different to at least one of the others for the week 11 (see the first row).
    \end{tablenotes}
\end{threeparttable}
}
\end{table}

\begin{table}
\centering
\scalebox{0.8}{
 \begin{threeparttable}
\caption{$\chi^2$ significance test of prevalence between different  age groups of women and men}
\label{femhomtest}
\begin{tabular}{|l| c| c |c |c| c| c |c |c |c |c |c|}
\hline
\backslashbox{Age group}{Week}  & 11 & 12 & 13 & 14 & 15 & 16 & 17 & 18 & 19 & 20 & 21\\ 
\hline
W-15 vs 5 others  &  \cellcolor{red} &  \cellcolor{green} & \cellcolor{green} & \cellcolor{green} & \cellcolor{green} & \cellcolor{green} & \cellcolor{green} & \cellcolor{green} & \cellcolor{red} & \cellcolor{red} & \cellcolor{red}\\
\hline
W15-44 vs 5 others &  \cellcolor{red} &  \cellcolor{green} & \cellcolor{green} & \cellcolor{red} & \cellcolor{red} & \cellcolor{red} & \cellcolor{red} & \cellcolor{red} & \cellcolor{red} & \cellcolor{red} & \cellcolor{red}\\
\hline
W45-64 vs 5 others &  \cellcolor{red} &  \cellcolor{red} & \cellcolor{red} & \cellcolor{red} & \cellcolor{red} & \cellcolor{red} & \cellcolor{red} & \cellcolor{red} & \cellcolor{red} & \cellcolor{red} & \cellcolor{red}\\ 
\hline
W65-74 vs 5 others  & \cellcolor{red} &  \cellcolor{red} & \cellcolor{green} & \cellcolor{red} & \cellcolor{red} & \cellcolor{red} & \cellcolor{red} & \cellcolor{red} & \cellcolor{red} & \cellcolor{red} & \cellcolor{red}\\
\hline
W75+ vs 5 others  & \cellcolor{red} &  \cellcolor{red} & \cellcolor{red} & \cellcolor{red} & \cellcolor{red} & \cellcolor{red} & \cellcolor{red} & \cellcolor{red} & \cellcolor{red} & \cellcolor{green} & \cellcolor{green}\\
\hline
\end{tabular}
   \begin{tablenotes}
   		 \item W-15, W15-44, W45-64, W65-74, W75+ : Female under 15, between 15-44, 45-64, 65-74 and 75 years and over.
        \item M-15, M15-44, M45-64, M65-74, M75+ : Male under 15, between 15-44, 45-64, 65-74 and 75 years and over.
        \item 5 others : M-15, M15-44, M45-64, M65-74, M75+.
         \item \colorbox{green}{ green week 21}: means W75+ has a a prevalence of Covid-19 significantly different to the ones of all the others for the week 21 (see the last row).
         \item \colorbox{red}{ red week 11}: means M-15 has prevalence of Covid-19 that is not significantly different to at least one of the others for the week 11 (see the first row).
    \end{tablenotes}
\end{threeparttable}
}
\end{table}

\begin{table}
\centering
\scalebox{0.8}{
 \begin{threeparttable}
\caption{The number of people tested, the number of people tested positive and the risk of Covid-19 for each age group}\label{agetable}
\begin{tabular}{|l| c| c c c c c c c c c c c|}
\hline
\backslashbox{people age group}{Week} &  & 11 & 12 & 13 & 14 & 15 & 16 & 17 & 18 & 19 & 20 & 21\\ 
\hline
\multirow{3}{*}{people under 15 years}  & pos & 3 &  11 & 35 & 57 & 33 & 17 & 23 & 13 & 11 & 33 & 14\\
						& tot & 46 & 189 & 315 & 419 & 452 & 472 & 513 & 429 & 798 & 2164 & 2175\\
						& risk & 6.52 & 5.82 & 11.11 & 13.6 & 7.3 & 3.6 & 4.48 & 3.03 & 1.38 & 1.52 & 0.64\\
\hline
\multirow{3}{*}{people between 15-44 years} & pos & 82 & 864 & 2693 & 2851 & 2148 & 1221 & 805 & 397 & 296 & 228 & 134\\
						& tot &  442 & 3864 & 11722 & 16613 & 16952 & 16340 & 16452 & 11494 & 13684 & 18804 & 14629\\
						& risk & 18.55 & 22.36 & 22.97 & 17.16 & 12.67 & 7.47 & 4.89 & 3.45 & 2.16 & 1.21 & 0.92\\
\hline
\multirow{3}{*}{people between 45-64 years} & pos & 75 & 924 & 2847 & 3015 & 2325 & 1249 & 742 & 355 & 237 & 197 & 92\\
						& tot & 320 & 2930 & 9819 & 14187 & 15127 & 14223 & 14405 & 10192 & 12051 & 15754 & 12231\\
						& risk & 23.44 & 31.54 & 28.99 & 21.25 & 15.37 & 8.78 & 5.15 & 3.48 & 1.97 & 1.25 & 0.75\\
\hline
\multirow{3}{*}{people between 65-74 years} & pos & 29 & 317 & 917 & 867 & 693 & 426 & 243 & 85 & 95 & 54 & 44\\
						& tot & 101 & 796 & 2710 & 3604 & 4082 & 3950 & 4170 & 2870 & 3555 & 5467 & 4624\\
						& risk & 28.71 & 39.82 & 33.84 & 24.06 & 16.98 & 10.78 & 5.83 & 2.96 & 2.67 & 0.99 & 0.95\\
\hline
\multirow{3}{*}{people 75 years and over} & pos & 32 & 429 & 1436 & 2091 & 3257 & 2182 & 1275 & 470 & 374 & 355 & 196\\
						& tot & 125 & 1449 & 4508 & 6199 & 12786 & 16806 & 16782 & 10888 & 10756 & 10557 & 7054\\
						& risk & 25.6 & 29.61 & 31.85 & 33.73 & 25.47 & 12.98 & 7.6 & 4.32 & 3.48 & 3.36 & 2.78\\
\hline
\end{tabular}
   \begin{tablenotes}
        \item pos : number of positive people.
        \item tot : total number of people tested.
        \item risk : Covid-19 prevalence rate ($\%$).
    \end{tablenotes}
\end{threeparttable}
}
\end{table}

The $\chi^2$ tests on the new partition of the population into $5$ age groups show that from week $14$  to week $16$ all the risks of the  subgroup are statistically different two-by-two (see Table~\ref{agetest}). This table has the same format as Tables~\ref{femtest}, \ref{homtest} and \ref{femhomtest}. 

\begin{table}
\centering
\scalebox{0.8}{
 \begin{threeparttable}
\caption{$\chi^2$ significance test of prevalence between different age groups}
\label{agetest}
\begin{tabular}{|l| c| c |c |c| c| c |c |c |c |c |c|}
\hline
\backslashbox{ Age group}{Week}  & 11 & 12 & 13 & 14 & 15 & 16 & 17 & 18 & 19 & 20 & 21\\ 
\hline
-15 vs 4 others  &  \cellcolor{green} &  \cellcolor{green} & \cellcolor{green} & \cellcolor{green} & \cellcolor{green} & \cellcolor{green} & \cellcolor{red} & \cellcolor{red} & \cellcolor{red} & \cellcolor{red} & \cellcolor{red}\\
\hline
15-44 vs 3 others &  \cellcolor{red} &  \cellcolor{green} & \cellcolor{green} & \cellcolor{green} & \cellcolor{green} & \cellcolor{green} & \cellcolor{red} & \cellcolor{red} & \cellcolor{red} & \cellcolor{red} & \cellcolor{red}\\
\hline
45-64 vs 2 others &  \cellcolor{red} &  \cellcolor{red} & \cellcolor{green} & \cellcolor{green} & \cellcolor{green} & \cellcolor{green} & \cellcolor{red} & \cellcolor{red} & \cellcolor{red} & \cellcolor{red} & \cellcolor{red}\\ 
\hline
65-74 vs 75+  & \cellcolor{red} &  \cellcolor{red} & \cellcolor{red} & \cellcolor{green} & \cellcolor{green} & \cellcolor{green} & \cellcolor{green} & \cellcolor{green} & \cellcolor{green} & \cellcolor{green} & \cellcolor{green}\\
\hline
\end{tabular}
   \begin{tablenotes}
   		 \item -15, 15-44, 45-64, 65-74, +75 : People under 15, between 15-44, 45-64, 65-74 and 75 years and over.
        \item 4 others : 15-44, 45-64, 65-74, 75+.
        \item 3 others : 45-64, 65-74, 75+.
        \item 2 others : 65-74, 75+.
        \item \colorbox{green}{ green week 21}: means people between 65-74 years have a prevalence of Covid-19 significantly different to the ones of 75 years and over people for the week 21 (see the last row).
        \item \colorbox{red}{ red week 17}: means people under 15  years have a prevalence of Covid-19 that is not significantly different to at least one of the others for the week 17 (see the first row).
        
    \end{tablenotes}
\end{threeparttable}
}
\end{table}

To generate an instance, we considered the daily data related to Covid-19 screening tests in the department of North in France. We disaggregated the data to obtain individual data. We assumed that the prevalences available on week $W$ are those computed for week $W-1$. Thus, we assigned to each subject the risk of the previous week $W-1$ according to his/her age group. We obtained six instances representing one-day Covid-19 screening tests from week $14$ to $16$. They include $54$, $54$, $157$, $146$, $104$ and $100$ subjects tested on March 30 and 31, and April 8, 9, 14 and 17 respectively. As these instances are associated with different weeks, they present a certain diversity, which allows us to evaluate the effect of the size and composition of an instance on the optimal solution. The aggregated data of the six instances are presented in Tables~\ref{Inst1} to \ref{Inst6}. For each instance, we report  the Covid-19 prevalence rate (risk), the total number of people tested and the number of positive ones by age group.

We set the test sensitivity of the RT-PCR tests $Se$ to $0.7$ and specificity $Sp$ to $0.95$. According to \cite{watson2020}, these are conservative values from systematic reviews \cite{arevalo2020}. $\lambda$ is set to $0.8$ to favor the minimization of expected false negatives.

The testing budget $B$ or the number of available tests to be conducted should clearly be at most $|S|-1$ to need partitioning $\mathcal S$, otherwise we test all the population individually.
To tighten $B$, starting from $B=|\mathcal S|-1$, we decrease the budget of available tests as much as we can until the algorithm fails to find a solution.

Note that another way to have an approximate value of $B$ is to simply compute $B' = \mathbb{E}[T(\Omega_1{}_{-(N+1)})]$ where $\Omega_1{}_{-(N+1)} = \{\mathcal S_1, \mathcal S_2, \dots, \mathcal S_N\}$. $B'$ can be used as starting value to decrease from instead of $B=|\mathcal S|-1$.

\subsection{Computational results}

We present the results of our experiments on real instances arising from Covid-19 screening tests performed in laboratories in the Northern department of France. Results are computed with the procedure proposed in Section~\ref{sec:algo}. The procedure is coded in C++ and experiments are run on an Intel Core i7-9850H CPU @ 2.60GHz 2.59GHz computer with 16 Gb of RAM. 

Researchers have shown the accuracy of \gt when the size of groups does not exceed a specific number. Using the standard COVID-19 RT-qPCR test, \cite{yelin2020} showed that a single positive subject can be detected in a pool of up to 32 persons, while \cite{benami2020} has shown the accuracy of limiting the group sizes to only eight subjects. We used both limits for the group size in our experiments.

The computational results on the six instances are reported in Tables~\ref{6instResults} and \ref{6instResults2}. Each row in the tables is associated with one instance. In the first three columns, we indicate the number of subjects, the minimum and the maximum  Covid-19 risk values. The fourth column reports the minimum number of tests needed. The optimal value of the weighted sum of the expected number of false-negative and false-positive classifications is reported in the fifth column. The sixth column reports the number of groups formed. The seventh and eighth columns report the minimum and maximum numbers of subjects in a group. The ninth column shows the expected gain in percentage in terms of test budget compared with the individual screening solution. The last column reports the computation time in seconds.

When we compare the results for a group size fixed to $8$ and $32$ in Table~\ref{6instResults}, we notice that for all instances, the number of tests required found when the group size i fixed to $8$ is greater or equal to that found when the group size is $32$. Nevertheless, the group size of $8$  tends to perform slightly better in terms of misclassification errors as measured by the objective function value (see column \textbf{objVal} in Table~\ref{6instResults}). A smaller objective function value corresponds to a higher test precision. However, when the population size increases, better results are obtained in terms of test budget and precision with a group size fixed to $32$. For example, for Instance 3 with $157$ subjects tested, the test budget and the test precision are $103$ and $12.80$ respectively when the group size is $32$. These values are $104$ and $12.91$ respectively when the group size is $8$.

The use of  \gt is always better than the current practice of testing each subject individually regardless of whether the group size is limited to $32$ or $8$. In our experiments on real instances, we can achieve more than $34\%$ of expected gain compared to the current practice (see column \textbf{Gain} in Table~\ref{6instResults}). This expected gain depends on the size of the population and the prevalence rate of the disease. We will have a much higher gain if we consider a large population with a low prevalence rate.

Concerning the optimal solution, we observe that the minimum number of tests required (\textbf{B}), the misclassification errors (\textbf{objVal}), the number of groups formed (\textbf{G}), the groups lower-bound (\textbf{minSubG}) and upper-bound (\textbf{maxSubG}) depend on the size of the population and the subject risks in an instance. For example, Instance 1 and 2 have the same number of subjects. However, the test precision is different in the two cases (see Table~\ref{6instResults}).

In Table~\ref{6instResults2}, we compare the results when the group size is fixed to $8$ and $32$ by setting the test budgets to the same value. We observe that a group size of $32$ leads to better results in terms of misclassification errors with large instances (see column \textbf{objVal} in Table~\ref{6instResults2}).

\begin{table}
\caption{\textbf{Instance 1:} 54 subjects resulting from Covid-19
screening tests on March 30, week 14, in city laboratories in the  department of North in France. The risks are those of week 13 for  the department of North.}
\centering
\label{Inst1}
\begin{tabular}{l c c c}
\hline
   Ages (years) & risk  & Number & Positive\\
\hline
-15 &  & 0 & 0\\
 15-44 & 0.238 & 18 & 1\\
 45-64 & 0.339 & 18 & 5 \\
 65-74 & 0.370 & 4 & 2\\
 75+ & 0.253 & 14 & 5\\
\textbf{Total} &  & \textbf{54}  & \textbf{13}\\
\hline
\end{tabular}
\end{table}

\begin{table}
\caption{\textbf{Instance 2:} 54 subjects resulting from Covid-19
screening tests on March 31, week 14, 
in city laboratories in the  department of North in France. The risks are those of week 13 for  the department of North.}
\centering
\label{Inst2}
\begin{tabular}{l c c c}
\hline
  Ages (years) & risk  & Number & Positive\\
\hline
-15 &  & 0 & 0\\
 15-44 & 0.238 & 18 & 3\\
 45-64 & 0.339 & 21 & 1 \\
 65-74 & 0.370 & 4 & 0\\
 75+ & 0.253 & 11 & 7\\
\textbf{Total} &  & \textbf{54}  & \textbf{11}\\
\hline
\end{tabular}
\end{table}

\begin{table}
\caption{\textbf{Instance 3:} 157 subjects resulting from Covid-19
screening tests on April 8, week 15, in city laboratories
in city laboratories in the  department of North in France. The risks are those of week 14 for  the department of North.}
\centering
\label{Inst3}
\begin{tabular}{l c c c}
\hline
   Ages (years) & risk  & Number & Positive\\
\hline
-15 & 0.000 & 2 & 0\\
 15-44 & 0.187 & 76 & 14\\
 45-64 & 0.152 & 44 & 6 \\
 65-74 & 0.125 & 13 & 5\\
 75+ & 0.369 & 22 & 7\\
\textbf{Total} &  & \textbf{157}  & \textbf{32}\\
\hline
\end{tabular}
\end{table}

\begin{table}
\caption{\textbf{Instance 4:} 146 subjects resulting from Covid-19
screening tests on April 9, week 15, in city laboratories 
in city laboratories in the  department of North in France. The risks are those of week 14 for  the department of North.}
\centering
\label{Inst4}
\begin{tabular}{l c c c}
\hline
  Ages (years) & risk  & Number & Positive\\
\hline
-15 & 0.000 & 1 & 0\\
 15-44 & 0.187 & 59 & 11\\
 45-64 & 0.152 & 45 & 8 \\
 65-74 & 0.125 & 12 & 2\\
 75+ & 0.369 & 29 & 7\\
\textbf{Total} &  & \textbf{146}  & \textbf{28}\\
\hline
\end{tabular}
\end{table}

\begin{table}
\caption{\textbf{Instance 5:} 104 subjects resulting from Covid-19
screening tests on April 14, week 16, in city laboratories 
in city laboratories in the  department of North in France. The risks are those of week 15 for  the department of North.}
\centering
\label{Inst5}
\begin{tabular}{l c c c}
\hline
  Ages (years) & risk  & Number & Positive\\
\hline
-15 & 0.000 & 3 & 1\\
 15-44 & 0.157 & 49 & 9\\
 45-64 & 0.199 & 34 & 9 \\
 65-74 & 0.176 & 4 & 0\\
 75+ & 0.277 & 14 & 1\\
\textbf{Total} &  & \textbf{104}  & \textbf{20}\\
\hline
\end{tabular}
\end{table}

\begin{table}
\caption{\textbf{Instance 6:} 100 subjects resulting from Covid-19
screening tests on April 17, week 16, 
in city laboratories in the  department of North in France. The risks are those of week 13 for  the department of North.}
\centering
\label{Inst6}
\begin{tabular}{l c c c}
\hline
  Ages (years) & risk  & Number & Positive\\
\hline
-15 & 0.000 & 0 & 0\\
 15-44 & 0.157 & 49 & 7\\
 45-64 & 0.199 & 36 & 5 \\
 65-74 & 0.176 & 9 & 0\\
 75+ & 0.277 & 6 & 1\\
\textbf{Total} &  & \textbf{100}  & \textbf{13}\\
\hline
\end{tabular}
\end{table}

\begin{table}
\centering
\scalebox{0.7}{
 \begin{threeparttable}
\caption{Results over the $6$ real instances with a limit of $8$ vs $32$ on the group size}
\label{6instResults}
\begin{tabular}{|l| c| c| c| c c| c c |c c|c c| c c| c c|c |}
\hline
\backslashbox{Instances }{}  & nbSub & minRisk & maxRisk & \multicolumn{2}{c|}{B} & \multicolumn{2}{c|}{objVal} &\multicolumn{2}{c|}{G} & \multicolumn{2}{c|}{ minSubG }& \multicolumn{2}{c|}{maxSubG} & \multicolumn{2}{c|}{Expected Gain ($\%$)}& CPU (s) \\ 
 &   &   &  & 8 &32 & 8 &32 & 8 & 32 & 8 & 32 & 8 & 32 & 8 & 32 & \\
\hline
Inst1 & 54  &  0.238 & 0.370  &  42 & 40 & 6.29 & 6.44 & 14 & 8 & 1 & 1 & 8 & 32 & 22.22 & 25.93 & 0.00 \\
\hline
Inst2 &  54 &  0.238 & 0.370 & 42 & 40 & 6.46 & 6.54 & 13 & 8 & 1 & 1 & 8 & 32 & 22.22 & 25.93 & 0.00 \\
\hline
Inst3 & 157  & 0.000 & 0.369 & 104 & 103 & 12.91 & 12.80 & 38 & 37 & 1 & 1 & 8 & 19 & 33.76 & 34.39 & 0.00 \\ 
\hline
Inst4 & 146  & 0.000 & 0.369  & 98 & 96 & 12.66 & 12.66 & 34 & 32 & 1 & 1 & 8 & 28 & 32.88 & 34.25 & 0.00 \\
\hline
Inst5 & 104  & 0.000 &  0.277  & 69 & 68 & 7.88 & 8.08 & 29 & 23 & 1 & 4 & 4 & 14 & 33.65 & 34.62 & 0.00 \\
\hline
Inst6 & 100  & 0.000 &  0.277  & 66 & 66 & 7.58 & 7.58 & 26 & 26 & 1 & 1 & 5 & 5 & 34.0 & 34.0 & 0.00 \\
\hline

\end{tabular}
   \begin{tablenotes}
   		 \item \textbf{nbSub:} number of subjects in the instance
   		 \item \textbf{minRisk:} minimum value of Covid-19 risk for the instance
   		 \item \textbf{maxRisk:} maximum value of Covid-19 risk for the instance
   		 \item \textbf{B:} the minimal number of available tests required to find a partition by the algorithm
   		 \item \textbf{objVal:} objective function value
   		 \item \textbf{G:} number of groups in the optimal solution
   		 \item \textbf{minSubG:} minimal number of subjects in a group
   		 \item \textbf{maxSubG:} maximal number of subjects in a group
   		 \item \textbf{CPU (s):} CPU time in seconds
   		 \item \textbf{Expected Gain ($\%$):} expected gain in percentage compared to the individual screening solution
    \end{tablenotes}
\end{threeparttable}
}
\end{table}

\begin{table}
\centering
\scalebox{0.7}{
 \begin{threeparttable}
\caption{Results over the $6$ real instances with a limit of $8$ vs $32$ on the group size and the same budget}
\label{6instResults2}
\begin{tabular}{|l| c| c| c| c c| c c |c c|c c| c c| c c|c |}
\hline
\backslashbox{Instances }{}  & nbSub & minRisk & maxRisk & \multicolumn{2}{c|}{B} & \multicolumn{2}{c|}{objVal} &\multicolumn{2}{c|}{G} & \multicolumn{2}{c|}{ minSubG }& \multicolumn{2}{c|}{maxSubG} & \multicolumn{2}{c|}{Expected Gain ($\%$)}& CPU (s) \\ 
 &   &   &  & 8 &32 & 8 &32 & 8 & 32 & 8 & 32 & 8 & 32 & 8 & 32 & \\
\hline
Inst1 & 54  &  0.238 & 0.370  &  42 & 42 & 6.29 & 6.00 & 14 & 14 & 1 & 1 & 8 & 29 & 22.22 & 22.22 & 0.00 \\
\hline
Inst2 &  54 &  0.238 & 0.370 &  42 & 42 & 6.46 & 6.10 & 13 & 14 & 1 & 1 & 4 & 4 & 9.25 & 9.25 & 0.00 \\
\hline
Inst3 & 157  & 0.000 & 0.369 & 104 & 104 & 12.91 & 12.62 & 38 & 40 & 1 & 1 & 8 & 16 & 33.76 & 33.76 & 0.00 \\ 
\hline
Inst4 & 146  & 0.000 & 0.369  & 98 & 98 & 12.66 & 12.24 & 34 & 38 & 1 & 1 & 8 & 21 & 32.88 & 32.88 & 0.00 \\
\hline
Inst5 & 104  & 0.000 &  0.277  & 69 & 69 & 7.88 & 7.88 & 29 & 29 & 1 & 1 & 4 & 4 & 33.65 & 33.65 & 0.00 \\
\hline
Inst6 & 100  & 0.000 &  0.277  & 66 & 66 & 7.58 & 7.58 & 26 & 26 & 1 & 1 & 5 & 5 & 34.0 & 34.0 & 0.00 \\
\hline
\end{tabular}
   \begin{tablenotes}
   		 \item \textbf{nbSub:} number of subjects in the instance
   		 \item \textbf{minRisk:} minimum value of Covid-19 risk for the instance
   		 \item \textbf{maxRisk:} maximum value of Covid-19 risk for the instance
   		 \item \textbf{B:} the minimal number of available tests required to find a partition by the algorithm
   		 \item \textbf{objVal:} objective function value
   		 \item \textbf{G:} number of groups in the optimal solution
   		 \item \textbf{minSubG:} minimal number of subjects in a group
   		 \item \textbf{maxSubG:} maximal number of subjects in a group
   		 \item \textbf{CPU (s):} CPU time in seconds
   		 \item \textbf{Expected Gain ($\%$):} expected gain in percentage compared to the individual screening solution
    \end{tablenotes}
\end{threeparttable}
}
\end{table}

\section{Conclusions}

\gt is a screening strategy that can be very efficient to test samples of a large population instead of individual screening since it reduces the required resources while maintaining the classification accuracy.  For a given maximum size of the groups to be identified, a key question is how to build such groups, which can be modeled as the Group Testing Design Problem (GTDP). In this paper, we extended the results of Aprahamian et al. \cite{aprahamian2018} when we impose a maximum size for the groups. 
Then, the \gtdp can still be modeled as a constrained shortest path problem, which can be solved in polynomial time due to the acyclic property of the underlying graph.  Computational results on instances derived from data provided by Sant\'e Publique France show that the proposed algorithm is able to provide solutions that reduce the number of tests by up to 34\% compared with an individual screening strategy while minimizing a convex combination of the expected numbers of false-negative and false-positive classifications in very short computation times.

The main perspective of this work is to investigate how the results could be further improved by designing a more relevant segmentation of the population. The population is currently segmented according to age groups set by Sant\'e Publique France, and the risks are calculated accordingly. We might benefit from designing a more relevant segmentation depending on the risks. Other relevant information, such as co-morbidities, could then be taken into account. Another perspective would be to consider additional constraints in the  \gtdp related to the technical characteristics of the clinical test, and to see how the properties, on which the approach is based, are still preserved.

\bibliographystyle{elsarticle-num}
\bibliography{biblioGroupTesting}

\end{document}